\documentclass[prd,twocolumn,nofootinbib,twocolumn]{revtex4-2}

\usepackage{graphicx, epsfig}
\usepackage{color}
\usepackage{mathrsfs}
\usepackage{bm}
\usepackage{amsmath,amssymb}
\usepackage{mathrsfs}
\usepackage[caption=false]{subfig}
\usepackage[normalem]{ulem}
\usepackage{mathtools}
\usepackage[x11names]{xcolor}

\usepackage{physics}
\usepackage{amsthm}
\usepackage{soul}
\usepackage[colorlinks = true, linkcolor = purple, urlcolor  = blue, citecolor = blue, anchorcolor = blue]{hyperref}
\usepackage{float}
\usepackage[caption=false]{subfig}
\graphicspath{{images/}}

\definecolor{fashionfuchsia}{rgb}{0.96, 0.0, 0.63}
\colorlet{no_so_fashion_purple}{blue!50!red}

\newcommand{\be}{\begin{equation}}
\newcommand{\ee}{\end{equation}}
\newcommand{\ba}{\begin{eqnarray}}
\newcommand{\ea}{\end{eqnarray}}

\newcommand{\nn}{\nonumber}

\newcommand{\la}{\langle}
\newcommand{\ra}{\rangle}

\newcommand{\bfZ}{{\bf Z}}

\newcommand{\bfOmega}{{\bf \Omega}}

\begin{document}

\hfill CERN-TH-2023-154

\title{Creating kinks with quantum mediation}
\author{Omer Albayrak$^*$, Tanmay Vachaspati$^{*,\dag}$}
\affiliation{
$^*$Physics Department, Arizona State University, Tempe,  Arizona 85287, USA.\\
$^\dag$Theoretical Physics Department, CERN, 1211 Geneva 23, Switzerland.
}

\begin{abstract}
We consider the creation of kink-antikink pairs of a scalar field $\phi$
by the scattering of classical wavepackets of a second scalar field, $\psi$, when
there are no direct interactions between $\phi$ and $\psi$. The creation becomes 
possible only due to a quantum field that interacts with both $\phi$ 
and $\psi$. We scan parameter space and find it favorable for kink production when the
initial wavepackets have large total energy and wide spatial extent but scatter at low 
velocities.
\end{abstract}

\maketitle

\section{Introduction}
\label{intro}

One of the most fascinating aspects of quantum field theory is the existence of 
non-perturbative topological structures (solitons) and their interactions with the 
perturbative excitations (particles) of the 
model~\cite{Rebbi:1985wg,Rajaraman:1982is,Vachaspati:2006zz,Manton:2004tk,Vilenkin:2000jqa}.
This area of research has received much attention but most of it has been relegated to treating the 
soliton sector as a fixed static or dynamical classical background. In any attempt in which 
solitons annihilate or are created, one is faced with the additional conceptual issue that
solitons are described in terms of classical fields while particles are quantum
excitations. A bridge between the soliton and particle sectors must also bridge between
classical and quantum behavior, unless one can overcome the difficult problem of treating
the soliton as a fully quantum object.

The creation of solitons by the scattering of particles~\cite{PhysRevLett.49.102,PhysRevD.71.025001,Romanczukiewicz:2005rm,Dutta:2008jt,PhysRevLett.105.081601,Demidov:2011dk,Demidov:2011eu,PhysRevD.87.065018,Dorey:2015sha,Demidov:2015nea} is of particular physical
interest. Sphalerons are classical solutions in the standard model that 
are intermediate states in baryon number violating processes that are necessary
to generate the cosmic matter-antimatter asymmetry~\cite{PhysRevD.30.2212,Morrissey:2012db}. 
If baryon number violation
is to be experimentally tested in particle accelerators, it will be necessary to
understand the creation of a sphaleron in particle collisions~\cite{papaefstathiou_phenomenology_2019}. 
The expectation
is that the process will be exponentially suppressed because perturbative
expansions are in powers of the coupling constant, while the sphaleron
and its interactions depend inversely on the coupling constant. 
Another process of interest is the production of magnetic monopoles 
in proton-proton or heavy ion scatterings such as at the Large Hadron Collider, a process that is being searched by the MoEDAL experiment~\cite{MoEDAL:2014ttp,MoEDAL:2016lxh,Mitsou:2022cuw}.
These searches are based on Schwinger pair production of magnetic 
monopoles~\cite{Affleck:1981ag,Affleck:1981bma}
in relatively strong magnetic fields that can be produced as heavy ions scatter in close
proximity~\cite{Ho:2019ads,Rajantie:2019oyc,Ho:2021uem}. The pair production
rate is evaluated using instanton methods.
Alternately, monopole creation by the scattering of large classical initial states of gauge
bosons has been considered in Ref.~\cite{Vachaspati:2016abz}.

While the two particle to soliton-antisoliton process is of interest because of
the way accelerators operate, one can envision situations where many 
particles may scatter and lead to the creation of solitons. This is the case
for baryon number violation at high temperatures such as in cosmology.
It may be possible that future particle machines may also involve $N$
particle scattering where $N$ can be large. Then the initial scattering
state may be described classically and the final state with solitons may
also be adequately described using classical physics. For example, we
may be interested in the production of magnetic monopoles in the scattering
of intense light. The problem in this setup is that the classical description
of light is given by Maxwell equations that are linear and, classically, light does not
interact with light. Colliding beams of intense light will simply pass
through each other in the classical description. Only when we include quantum 
effects such as box diagrams does light interact with light~\cite{PhysRev.44.855.2,PhysRev.46.1087,Euler:1935zz}. Such quantum
effects need only occur at intermediate stages in the scattering -- the
initial and final states can be described classically.

Guided by these motivations we have studied the creation of 1+1 
dimensional kinks in the scattering of classical initial states but those 
that interact with the
classical kink degrees of freedom only by a quantum ``bridge''. Then we 
have three fields: $\phi$ the classical field that has kink configurations, 
$\psi$ the classical field that defines the initial scattering state, and $\rho$
the quantum field that bridges between $\phi$ and $\psi$, the two classical
fields.

We will set up the field theory model in more detail in Sec.~\ref{model}. In Sec.~\ref{initial} 
we describe the kink solution and its energy along with the initial conditions for 
the model. We describe the numerical method in Sec.~\ref{numerical} and present the 
parameters that we have used. In Sec.~\ref{results}  we analyze few typical cases in 
detail and display the parameter space suitable for kink production. Finally we discuss 
our results in Sec.~\ref{conclusions}.

\section{Model}
\label{model}

The Lagrangian for the model we study is,
\ba
L &=&  \frac{1}{2} (\partial_\mu \phi )^2 - \frac{\lambda}{4} (\phi^2-\eta^2)^2
+ \frac{1}{2} (\partial_\mu \psi )^2 - \frac{m_\psi^2}{2} \psi^2
\nn \\ 
&+&
\frac{1}{2} (\partial_\mu \rho )^2 
- \frac{1}{2} \left ( m_\rho^2 + \alpha \phi^2 + \beta \psi^2 \right ) \rho^2
\ea
and the equations of motion are
\ba
    \Box\phi  + \lambda \phi (\phi^2 -\eta^2) +\alpha {\rho}^2\phi &=& 0 
    \label{eqn:phiBefore}\\
    \Box\psi  + m^{2}_{\psi}\psi +\beta {\rho}^2\psi &=& 0 \label{eqn:psiBefore} \\
     \Box\rho  +  \left ( m_\rho^2 + \alpha \phi^2 + \beta \psi^2 \right ) \rho &=& 0. \label{eqn:rhoBefore}
\ea
Ideally all three fields should be treated in quantum field theory but this may be
unnecessary for our purposes.
We will treat the incoming field, $\psi$, as a classical field. This can be justified
on the grounds that our incoming state has high occupation number, typically of
order $N \sim 10^3$. (Although high occupation number is not sufficient for a field to
behave classically, we will use this criterion as a guide.) We expect quantum 
corrections to the initial state to be suppressed by $1/N$. The initial state evolves
to excite quanta of the $\rho$ field. Thus the occupation number of the $\rho$
field also grows and parametrically it should go as $\beta^2 N$ where $\beta$
is the coupling constant for the interaction between $\psi$ and $\rho$. We will
take $\beta=0.5$, and so the occupation number of $\rho$ is of order $10^2$.
Finally, $\phi$ is treated classically. The occupation number for $\phi$ quanta
is expected to be of order $\alpha^2(\beta^2 N)$ where $\alpha$ denotes the
coupling between $\rho$ and $\phi$ and is taken to be 0.5. This means that the
occupation number of $\phi$ particles is also large, in the 10-100 range and quantum
corrections to the evolution of $\phi$ can be expected to be suppressed by
$(\alpha^2\beta^2 N)^{-1} \sim 0.1$.

The key approximation enters when we include quantum backreaction on the
classical background.
Since $\rho$ is a quantum operator that also appears in the $\phi$ and $\psi$ classical
equations of motion, we use the semiclassical approximation to write
\ba
    \Box\phi  + \lambda \phi (\phi^2 -\eta^2) +\alpha \la {\rho}^2 \ra \phi &=& 0 
    \label{eqn:phiSC}\\
    \Box\psi  + m^{2}_{\psi}\psi +\beta \la {\rho}^2 \ra \psi &=& 0 \label{eqn:psiSC}
\ea
where the expectation of $\rho^2$ is taken in its initial quantum state. (We work in
the Heisenberg representation in which operators evolve but the quantum states
do not.) The equation for the quantum operator $\rho$ can be solved since the
equation is linear in $\rho$. As discussed in Refs.~\cite{Vachaspati:2018pps,Vachaspati:2018llo,Vachaspati:2018hcu},
the solution is obtained using a ``classical-quantum correspondence'' (CQC) that 
we now summarize.

Starting with the action for the field $\rho(t,x)$;
\begin{equation}
    \mathcal{S}_{\rho} = \int d^2x \Bigg[ \frac{1}{2}(\partial_{\mu}\rho)^2 - \frac{1}{2} (m_{\rho}^2 + \alpha\phi^2 +\beta\psi^2) \rho^2 \Bigg]
    \label{rhoS}
\end{equation}
This action describes the massive quantum field $\rho$ in the time-dependent background of 
$\phi(t,x)$ and $\psi(t,x)$.  We continue with discretizing the action in space. On a lattice with 
$N$ sites with lattice spacing $a$, for any field, $f$, we define
\begin{equation}
    f(t,x) \rightarrow f(t,ja) = f_j(t)  \label{eqn:discretization} 
\end{equation}
\begin{equation}
    \nabla^2 f_j(t) =\frac{1}{a^2} (f_{j+1}(t)-2f_j(t)+f_{j-1}(t)).
    \label{eqn:finiteDifSch}
\end{equation}
where $j = 1,2,...,N$. The lattice under consideration is subjected to periodic boundary conditions 
such that for any field $f(t,x)$, $f_{j+N}(t) = f_j(t)$. The discretized action \eqref{rhoS} reads
\begin{align}
    \mathcal{S}_{\rho} = \int dt \frac{1}{a} \Bigg[ \frac{1}{2} \dot{\mathbf{x}}^T \dot{  \mathbf{x}} - \frac{1}{2} \mathbf{x}^T  \mathbf{\Omega}^2 \mathbf{x} \Bigg]
\end{align}
where $\mathbf{x}=(a\rho_{1},...,a\rho_{N})^T$ 
and $\mathbf{\Omega}^2$ is an $N\cross N$ matrix is given by
\begin{equation}
    \Omega^2_{jk} = \begin{cases} 
      2/a^2 + (m_{\rho}^2 + \alpha \phi_j^2 + \beta \psi_j^2)  & j=k \\
      -1/a^2 & j=k\pm 1 (\bmod N)\\
      0 & $otherwise.$ \label{Om2}
   \end{cases} 
\end{equation}
The mod $N$ 
is due to the periodic boundary conditions of the lattice. 
The energy of the system given by the action $\mathcal{S}_{\rho}$ as above can be derived as follows 
\begin{equation}
    H_{\rho} = \frac{a}{2}  \mathbf{p}^T\mathbf{p} 
    + \frac{1}{2a} \mathbf{x}^T  \boldsymbol{\Omega}^2 \mathbf{x}
    \label{hamrho}
\end{equation} 
where $\mathbf{p} = \mathbf{\dot{x}}/a$. 
This expression is precisely the Hamiltonian of $N$ coupled harmonic oscillators with time-dependent 
spring constant matrix.

The Hamiltonian in \eqref{hamrho} can be mapped to a classical system.
The technique is to use the Bogoliubov transformations to map the $N$ coupled quantum harmonic 
oscillator problem to an $N^2$ classical harmonic oscillator problem whose variables are
written as an $N\times N$ matrix $\mathbf{Z}(t)=[Z_{jk}(t)]$ and the corresponding momentum
matrix $\mathbf{P}(t)=[P_{jk}(t)] = \mathbf{\dot{Z}}/a$ ~\cite{Vachaspati:2018hcu}.
The mapping is given by
\begin{align}
     \mathbf{x}&=\mathbf{Z^*} \mathbf{a}_0 + \mathbf{Z}\mathbf{a}^{\dagger T}_0\label{eqn:xtoZ}\\
     \mathbf{p}&=\mathbf{P^*} \mathbf{a}_0 + \mathbf{P}\mathbf{a}^{\dagger T}_0\label{eqn:ptoZ}
\end{align}
where $\mathbf{a} = (a_{1},...,a_{N})^T$ and 
$\mathbf{a}^{\dagger} = (a_{1}^{\dagger},...,a_{N}^{\dagger})$ are the ladder operators 
for each of the $N$ harmonic oscillators and the subscript $"0"$ represents the operators
at the initial time $t_0$.  The quantum field $\rho(x,t)$ can now be represented 
in terms of corresponding expressions of $\mathbf{Z}$ and $\mathbf{P}$ 
(or equivalently $\mathbf{\dot{Z}}$) 
using \eqref{eqn:xtoZ} and \eqref{eqn:ptoZ}. 

The resulting classical system of $\mathbf{Z}(t)$ has the following action
\begin{equation}
    \mathcal{S}_c = \int dt \frac{1}{2a} \textrm{Tr }[\dot{\mathbf{Z}}^{\dagger}\dot{\mathbf{Z}}
    - \mathbf{Z}^{\dagger}\mathbf{\Omega}^2 \mathbf{Z}]
    \label{eqn:Z-action}
\end{equation}
and the equations of motion are
\begin{equation}
    \mathbf{\ddot{Z}} + \mathbf{\Omega}^2 \mathbf{Z} = 0. \label{eqn:Zeom}
\end{equation}
which are to be solved with the initial conditions,
\begin{equation}
    \mathbf{Z}_0 = -i \sqrt{\frac{a}{2}} \sqrt{\bf \Omega}^{-1} \quad \textrm{and} \quad  \mathbf{\dot{Z}}_0 = \sqrt{\frac{a}{2}} \sqrt{\bf \Omega}. \label{eqn:Z_ics}
\end{equation}
Since the CQC provides an exact correspondence of the quantum problem into its classical 
counterpart, from now on we only need equation (\ref{eqn:Zeom}) and initial conditions 
(\ref{eqn:Z_ics}) to fully understand the time evolution of the quantum field. 
The quantum evolution of $\rho$ is then obtained from \eqref{eqn:xtoZ} and \eqref{eqn:ptoZ}.

The vacuum expectation value of $\rho^2$
at the spatial point labelled by $i$ can be written in terms of $\bfZ$ as,
\be
\la \rho_i^2 \ra = \frac{1}{a^2} \sum_{j=1}^N Z_{ij}^* Z_{ij}.
\label{larho2ra}
\ee
using \eqref{eqn:xtoZ}. Therefore, from \eqref{eqn:phiSC} and \eqref{eqn:psiSC}, the discretized equations we 
would like to solve for $\phi$ and $\psi$ are
\ba
&& \hskip -0.75 cm
    {\ddot \phi_i} - \nabla^2\phi_i + \lambda \phi_i (\phi_i^2 -\eta^2) 
                   +\frac{\alpha }{a^2} \sum_{j=1}^N Z_{ij}^* Z_{ij}  \phi_i = 0 
    \label{eqn:phiSCdisc}\\
 &&
    {\ddot \psi_i} - \nabla^2\psi_i + m^2_{\psi}\psi_i 
    +\frac{\beta }{a^2} \sum_{j=1}^N Z_{ij}^* Z_{ij} \psi_i = 0 \label{eqn:psiSCdisc}
\ea
where we use second order spatial differences as in \eqref{eqn:finiteDifSch}
to calculate the Laplacians.
The equation for $Z_{ij}$ is
\be
{\ddot Z}_{ij} + \Omega^2_{ik} Z_{kj} = 0.
\label{eqn:Zij}
\ee

The system of equations \eqref{eqn:phiSCdisc}, \eqref{eqn:psiSCdisc} and
\eqref{eqn:Zij} need to be solved with suitable boundary conditions that we
will discuss below. Before proceeding to the solution, however, the issue
of renormalization needs to be addressed.

The parameters appearing in the above equations of motion are bare parameters 
that will get dressed by quantum effects. This can also be seen by realizing that
the quantity $\la \rho^2_i \ra$ in \textcolor{red}{\eqref{larho2ra}} diverges as $\log(N)$ as
$N \to \infty$. The divergence can be absorbed in the mass  parameters 
$m_{\phi}$ and $m_{\psi}$~\cite{Mukhopadhyay:2021wmu,Mukhopadhyay:2023zmc}.
Equivalently, we can subtract out the fluctuations in the trivial vacuum,
\be
\la \rho_i^2 \ra \to \la \rho_i^2 \ra - \la \rho_i^2 \ra_0
\ee
where, 
\be
\la \rho_i^2 \ra_0 \equiv  \frac{1}{a^2} \sum_{j=1}^N Z_{ij}^* Z_{ij} \biggr |_0.
\label{larho2raren}
\ee
The ``0'' subscript refers to the trivial vacuum with $\phi=\eta$ and $\psi=0$.

The energy in the quantum field $\rho$ can now be written as
\begin{equation}
    E_{\rho} = \frac{1}{2a} \textrm{Tr }[\mathbf{\dot{Z}}^{\dagger}\mathbf{\dot{Z}} + \mathbf{Z}^{\dagger}\mathbf{\Omega}^2\mathbf{Z}]. \label{eqn:energy_rho}
\end{equation}
and the discrete energy density,
\begin{align}
    \epsilon_{\rho,i} &= \frac{1}{a^2} \sum_k  \biggl \{ \frac{1}{2} |\dot{Z}_{ij}|^2 + \frac{1}{4a^2}  \biggl [|Z_{i+1j}-Z_{ij}|^2  \\ 
    &+ |Z_{ij}-Z_{i-1j}|^2   \biggl] 
     + \frac{1}{2}\left[ m_{\rho}^2 + \alpha\phi_j^2 + \beta\psi_j^2 \right] |Z_{ij}|^2 \biggl \} \nonumber.
\end{align}
Owing to the last term, this expression also suffers from the divergence mentioned above. 
We use the same renormalizing scheme to remove the lattice dependence and to obtain
a finite expression even as $N \to \infty$,
\begin{equation}
    \epsilon_{\rho,i}^{R} =  \epsilon_{\rho,i}-  \frac{1}{2}\left[ m_{\rho}^2 + \alpha\phi_i^2 + \beta\psi_i^2 \right]\left\langle \hat{\rho}^2_i\right\rangle_0 - \epsilon_{\rho,i}\vert_{_0}
\end{equation}
where the last term is added for the purpose of subtracting out the zero-point energy. 
The total energy of $\rho$ is similarly defined,
\be
    E_{\rho}^{R} =  E_{\rho} -  \frac{1}{2} \sum_{i=1}^N \left[ m_{\rho}^2 + \alpha\phi_i^2 + \beta\psi_i^2 \right]\left\langle \hat{\rho}^2_i\right\rangle_0 - E_{\rho}\vert_{_0}.
\ee
By adding the energy of the fields $\phi$ and $\psi$ the total conserved energy of the system is,
\begin{equation}
    E = E_{\phi+\psi} + E_{\rho}^R  
\end{equation}
where $E_{\phi+\psi}$ defined as
\begin{align}
     E_{\phi+\psi} &= \sum_i \biggl \{ \frac{1}{2}\left[\dot{\phi}_i^2 + \phi_{i}'^2+ \dot{\psi}_i^2 + \psi_{i}'^2 \right]\nonumber\\ 
     & \hskip 1.5 cm
     + \frac{1}{2}m_{\psi}^2 \psi_{i}^2 +\frac{\lambda}{4}\left(\phi_i^2-\eta^2\right)^2 \biggl \} . 
     \label{Ephipsi}
\end{align}
Spatial first derivatives are calculated using central differencing,
\be
f_i' = \frac{f_{i+1}-f_{i-1}}{2a}.
\ee

To summarize this section, the final equations we wish to solve are, 
\ba
&&
    {\ddot \phi_i} - \nabla^2\phi_i + \lambda \phi_i (\phi_i^2 -\eta^2) \nn \\
    && \hskip 1 cm
                   +\frac{\alpha }{a^2} \sum_{j=1}^N \left ( Z_{ij}^* Z_{ij}  - Z_{ij}^* Z_{ij}\biggr |_0 \right ) \phi_i = 0 
    \label{eqn:phiren}\\
 &&
    {\ddot \psi_i} - \nabla^2\psi_i + m^2_{\psi}\psi_i  \nn \\
    &&  \hskip 1 cm
    +\frac{\beta }{a^2} \sum_{j=1}^N  \left ( Z_{ij}^* Z_{ij}  - Z_{ij}^* Z_{ij}\biggr |_0 \right ) \psi_i = 0 \label{eqn:psiren}
\ea
and also Eq.~\eqref{eqn:Zij} for $\bfZ$.

\section{Initial conditions}
\label{initial}

We are interested in the creation of $Z_2$ kinks of $\phi$ due to collisions of classical 
wavepackets of $\psi$. 
The kink configurations are solutions of the model,
\be
   L_{\phi} =  \frac{1}{2} (\partial_\mu \phi )^2 - \frac{\lambda}{4} (\phi^2-\eta^2)^2
\ee
and boosted kinks are given by the solutions,
\be
    \phi_K(t,x) = \pm \eta \tanh \left ( \sqrt{\frac{\lambda}{2}} \ \eta \gamma (x-vt) \right )
\ee
where the Lorentz boost factor $\gamma=1/\sqrt{1-v^2}$, the $+$ sign denotes
a kink, and a $-$ sign an antikink. The energy of the kink (or antikink) is given by
\be
    E_K = \gamma \ \frac{2\sqrt{2}}{3} \sqrt{\lambda} \eta^3.
    \label{EK}
\ee

\begin{figure}
\includegraphics[width=0.47\textwidth,angle=0]{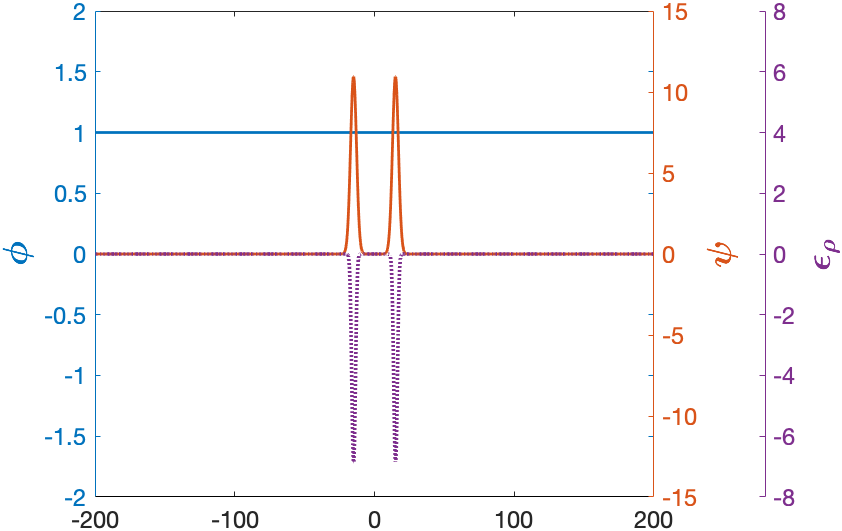}
\caption{Initial field configurations for $\phi$ (blue) and $\psi$ (orange)
and initial renormalized energy density of $\rho$ (dashed purple). 
}
\label{figinitial}
\end{figure}

In the classical scattering of $\psi$, kinks cannot be created without the participation of
the quantum field $\rho$ because $\phi$ and
$\psi$ have no direct coupling. Initially there are no kinks and we take $\phi$ to be in its
vacuum state,
\be
\phi(t=0,x)=\eta , \ \ {\dot \phi}(t=0,x)=0.
\ee
Our choice for the initial conditions for $\psi$ contains two Gaussian wavepackets 
that move towards each other with velocity $v$. Then,
\be
\psi(t=0,x) = F(\gamma (x+x_0)) + F(\gamma (x-x_0))
\label{psit0}
\ee
where $\gamma = 1/\sqrt{1-v^2}$, $2x_0$ is the initial ($t=0$) wavepacket separation, and
\be
F(x)= A e^{-kx^2}.
\ee
We also have,
\be
{\dot \psi}(t=0,x) = \gamma v \left [ F'(\gamma(x+x_0)) - F'(\gamma(x-x_0)) \right ]
\label{dotpsit0}
\ee
where primes denote derivatives with respect to the argument.

The quantum field $\rho$ is initially assumed to be in its ground state in the background
of $\phi(t=0,x)$ and $\psi(t=0,x)$. In terms of $\bfZ$ this is given by Eq.~\eqref{eqn:Z_ics}
where the matrix $\bfOmega^2_0$ is evaluated from \eqref{Om2} using the initial
values of $\phi$ and $\psi$.

In Fig.~\ref{figinitial} we show the $\phi$ and $\psi$ fields and the renormalized energy
density in $\rho$ at the initial time.

\section{Numerical method}
\label{numerical}

There are a large number of parameters that we need to fix before we can solve the
equations. We choose
\be
\lambda = 1, \ \ \eta = 1, \ \  \alpha = 0.5, \ \ m_{\rho} =1, \ \  m_\psi = 1, \ \ \beta = 0.5
\ee

The equations of motion are evolved using 
the position Verlet method with lattice spacing $a=0.4$ and
time step $dt=a/50$ on a periodic lattice with $N=1000$. The code is evolved
for less than a light-crossing time to prevent interference from excitations that
propagate all the way across the lattice.

There are also several parameters associated with the initial conditions: $x_0$, $A$, $v$ and $k$.
The initial separation of the Gaussian wavepackets is fixed to be $2x_0=30$. 
This is large enough that the overlap of the Gaussian wavepackets
is minimal for all runs.
We scan over $A$, $v$ and $k$ in the following intervals,
\be
A \in [7,16],  \ \ v \in [0.1,0.8], \ \ k= 0.03, 0.1, 0.3.
\label{Avk}
\ee

There is some ambiguity in deciding if the scattering has led to kink-antikink
production. The simplest definition is to identify a zero of $\phi$ as a kink or
an antikink (depending on the gradient of $\phi$ at the location of the zero).
However, two zeros representing a kink and an antikink may be very close 
to each other and they may eventually annihilate. A further refinement of
the criterion that we adopt is to require that the distance between zeros be larger than 
four times the kink width and should increase with time.

\section{Results}
\label{results}

In this section, we present our simulation results based on the methods mentioned above. 
Initially, we analyze a few distinct cases as examples for kink creation. Subsequently, we 
study the regions within parameter space that satisfy the conditions necessary for kink formation.

\begin{figure*}
\centering
\subfloat[]{\includegraphics[width=0.32\textwidth]
        {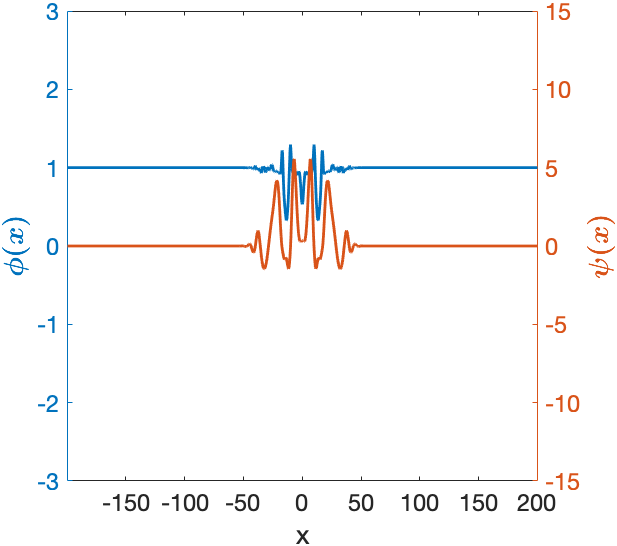}
        \label{fig:phi-psi-docile-scatter}}
\subfloat[]{\includegraphics[width=0.32\textwidth]
        {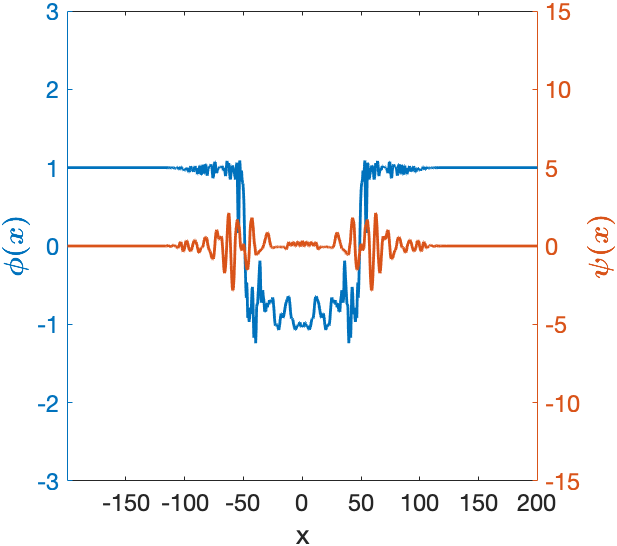}
        \label{fig:phi-psi-docile-formation}}
\subfloat[]{\includegraphics[width=0.32\textwidth]
        {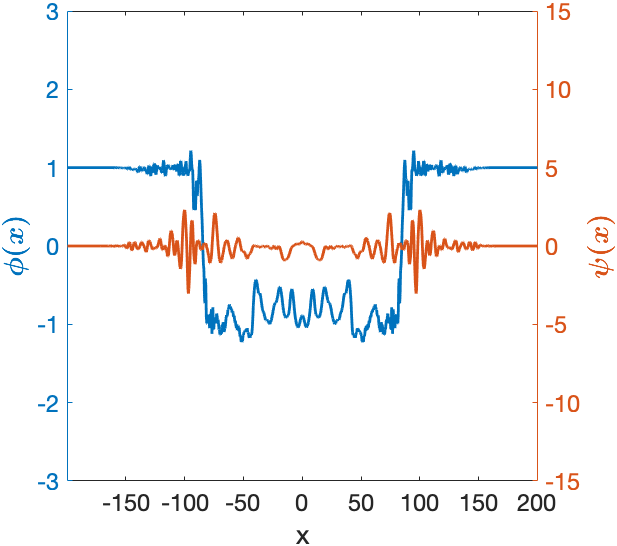}
        \label{fig:phi-psi-docile-movedaway}} \hfill
\caption{\label{fig:phi-psi-evol-docile-main} Three snapshots of the time evolution of the fields $\phi$ and $\psi$ with initial parameters $v=0.3$ and $A=11.0$. This is a clear case where a kink-antikink pair is produced. 
}
\end{figure*}

\begin{figure*}
    \centering
    \subfloat[]
    {\includegraphics[width=0.495\textwidth]
            {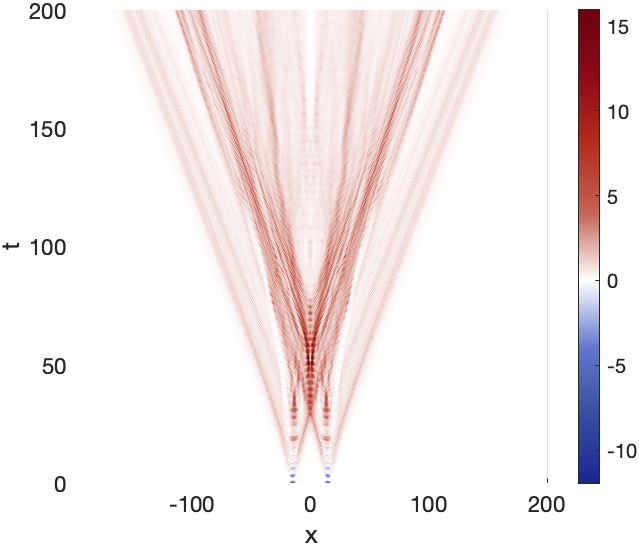}\hfill
            \label{fig:Z-ED-docile}}\hfill
    \subfloat[]
    {\includegraphics[width=0.45\textwidth]
        {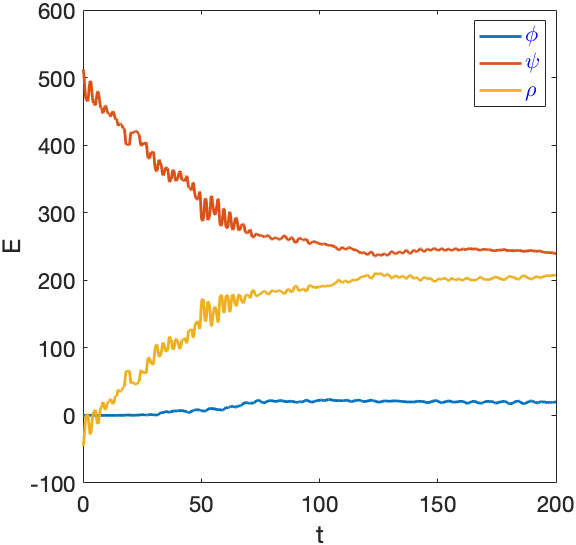}
        \label{fig:total-energy-docile}}\hfill
    \caption{\label{fig:energy-docile} 
    (a)  Evolution of the energy density $\epsilon_{\rho,i}$ for $v=0.3$ and $A=11.0$. At the initial time, 
    the quantum fluctuations
    are affected by the wavepackets of $\psi$ and there is non-vanishing energy density of $\rho$ within 
    the wavepackets. Once kinks are created ($t \gtrsim 50$), energy in the quantum fluctuations of $\rho$
    are carried by the kink-antikink pair. In addition, $\rho$ particles are radiated.
    (b) Total energies of the individual fields over time. For $\phi$ and $\psi$ only their kinetic, gradient and 
    potential terms are included (see \eqref{Ephipsi}). The suitably renormalized Interaction energy is included 
    in $\rho$. 
    The final energy in the kink field, $\phi$, is  $E_{\phi,{\rm final}} \sim 21.02$ which is about 10 times
    the energy in the kink-antikink pair.}
\end{figure*}

In Fig.~\ref{fig:phi-psi-evol-docile-main} we illustrate a clear case of kink production.
The three snapshots of the evolution for $v=0.3$ and $A=11.0$ show the collision
of the $\psi$ wavepackets and the creation of a kink-antikink pair that separates out
with velocities $\pm$0.68 respectively. Initially there is some energy in
$\rho$ that propagates together with the incoming wavepackets. After the collision,
if kinks are created, they too carry some $\rho$ energy along with them. 
In addition, we observe that there is energy in $\rho$ not directly related to the interactions 
with the initial Gaussian wavepackets or the final kinks. This energy is in the form of 
quantum radiation and can be seen on Fig~\ref{fig:Z-ED-docile}. 
The evolution of the total energy in the various fields
is shown in Fig.~\ref{fig:total-energy-docile}.

From Fig.~\ref{fig:total-energy-docile} we see that the initial energy is $\sim 500$
in units of $\sqrt{\lambda}$ whereas the energy of a kink $\sim 1$ from \eqref{EK} and that of a
kink-antikink pair is $\sim 2$. The collision has therefore converted less than a percent
of the initial energy into solitons; the rest is in radiative modes.

\begin{figure*}
\centering
\subfloat[] {\includegraphics[width=0.32\textwidth]
                {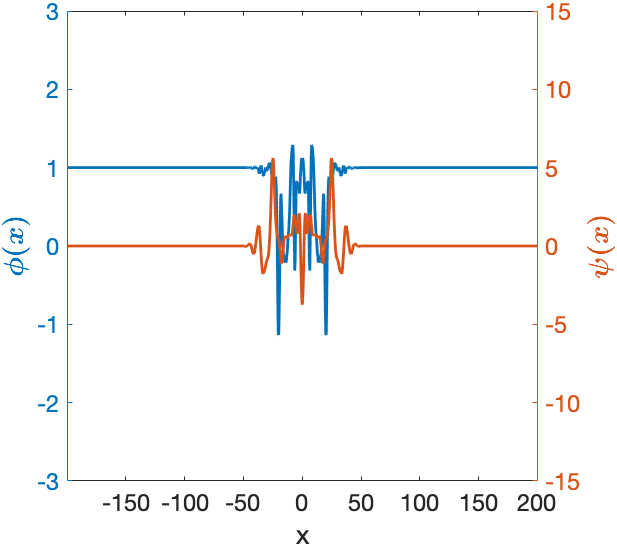}
                \label{fig:phi-psi-chaotic-scatter}}
\subfloat[] {\includegraphics[width=0.32\textwidth]
                {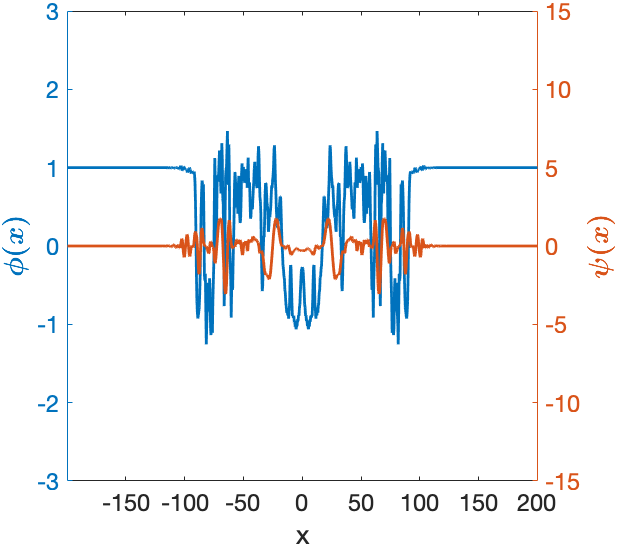}
                \label{fig:phi-psi-chaotic-formation}}
\subfloat[] {\includegraphics[width=0.32\textwidth]
                {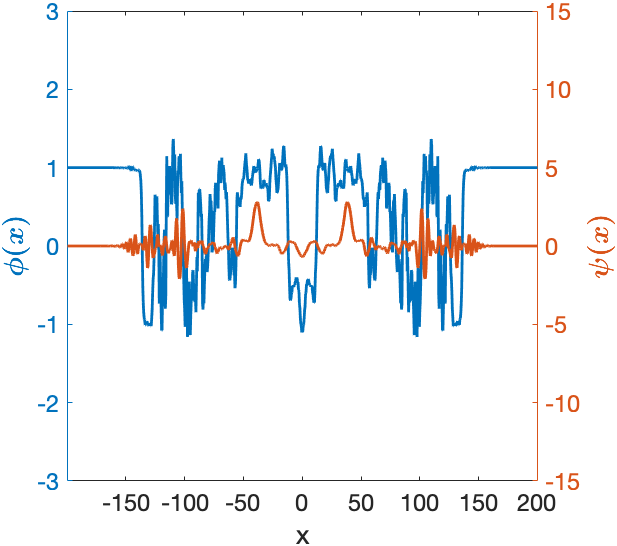}
                \label{fig:phi-psi-chaotic-movedaway}}
\caption{\label{fig:phi-psi-evol-chaotic-main} Three snapshots of the time evolution of the fields $\phi$ and $\psi$ with initial parameters $v=0.25$ and $A=13.5$. We observe a somewhat chaotic behaviour where there are multiple kink-antikink creation.
}
\end{figure*}

\begin{figure*}
    \centering
    \subfloat[]
    {\includegraphics[width=0.495\textwidth]
        {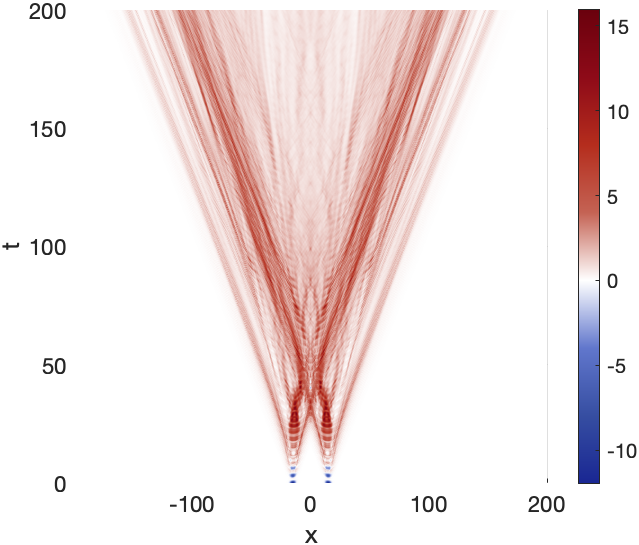}
        \label{fig:Z-ED-chaotic}}\hfill
    \subfloat[]
        {\includegraphics[width=0.45\textwidth]
        {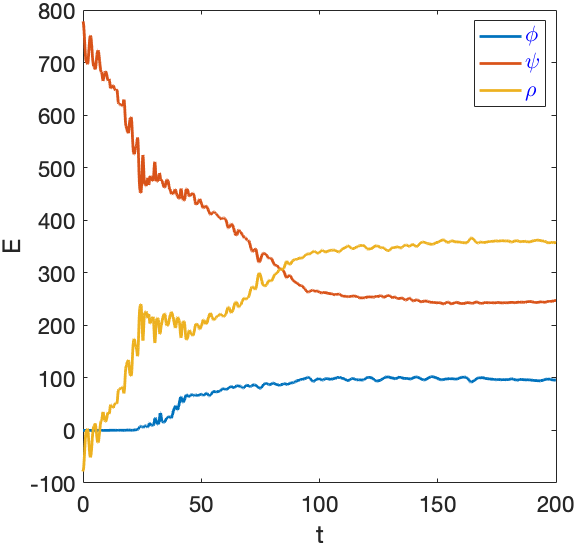}
        \label{fig:total-energy-chaotic}}\hfill
    \caption{\label{fig:energy-chaotic} 
    (a) Space-time plot of energy density $\epsilon_{\rho,i}$. Imprints of the initial $\psi$ wave packets and resulting kink-antikinks in $\phi$ are observed.
    (b) Energies of individual fields over time. For $\phi$ and $\psi$ only kinetic, gradient and potential terms are included. Interaction energy is included in $\rho$ with apt renormalization.
    $E_{\phi,{\rm final}} \sim 101.8 ~ E_{K}|_{\gamma=1}$.
    The parameters are $v=0.25$ and $A=13.5$.
    }
\end{figure*}

\begin{figure*}
    \centering
    \subfloat[]
    {\includegraphics[width=0.40\textwidth]
        {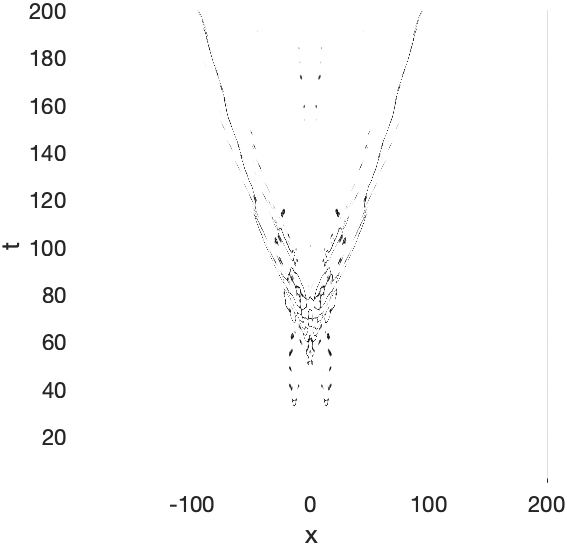}
        \label{fig:zeros-docile}}
        \hfill
    \subfloat[]
        {\includegraphics[width=0.40\textwidth]
        {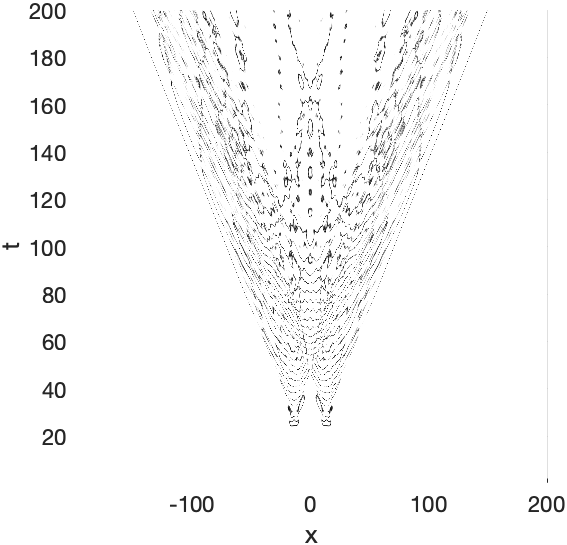}
        \label{fig:zeros-chaotic}}
        \hfill
    \caption{\label{fig:phi-zeros} The space-time graphs of zeros of $\phi$ for cases (a) $v=0.3$, $A=11.0$  and (b) $v=0.25$, $A=13.5$. 
 Only zeros that are well separated (four kink widths) and moving away with time from their neighbors are counted as kink-antikinks. These plots also display the kink-antikink pairs that are created but annihilate during the simulation.
    }
\end{figure*}

\begin{figure*}
    \centering
    \subfloat[]
    {\includegraphics[width=0.30\textwidth]
        {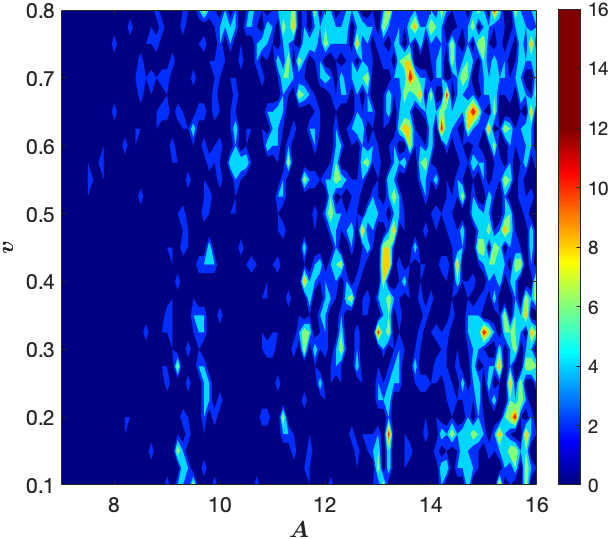}
        \label{fig:kinkSpect-thick}}\hfill
    \subfloat[]
    {\includegraphics[width=0.30\textwidth]
        {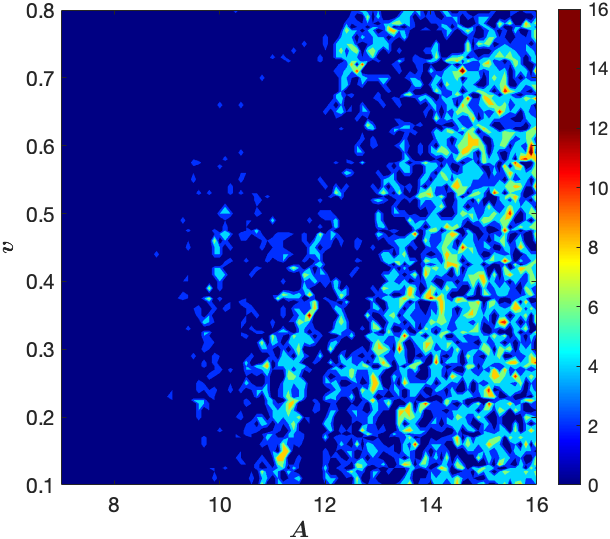}
        \label{fig:vAfork0p1}}\hfill
    \subfloat[]
        {\includegraphics[width=0.30\textwidth]
        {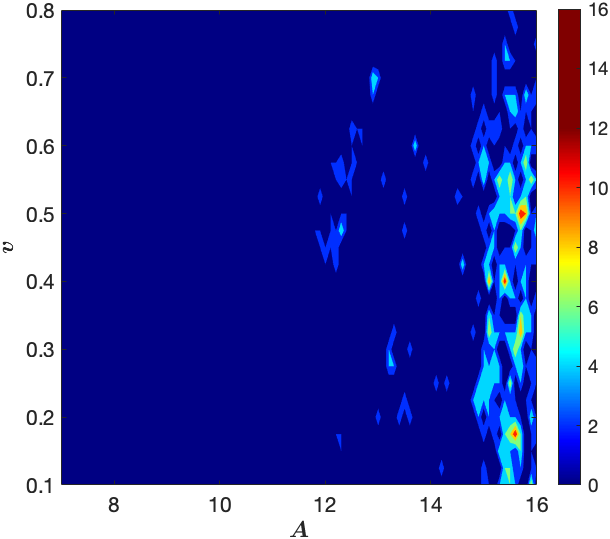}
        \label{fig:kinkSpect-thin}}\hfill
    \caption{\label{fig:kinkSpect} 
    Kink-antikink pair production in the amplitude ($A$) and velocity ($v$) plane as a contour plot for 
    (a) $k=0.03$, (b) $k=0.1$, and (c) $k=0.3$. 
    The color bar shows the total number of kink-antikinks produced -- a kink-antikink pair
    counts as 2 on the color bar.
    The gaps are genuine and show chaotic behavior  -- simply increasing
    the amplitude, for example, does not guarantee kink production. 
    Since the width of the initial wavepackets decreases as
    $k$ increases, the plots show that kink production is favored for larger widths.
    }
\end{figure*}

\begin{figure*}
    \centering
    \subfloat[]
    {\includegraphics[width=0.30\textwidth]
        {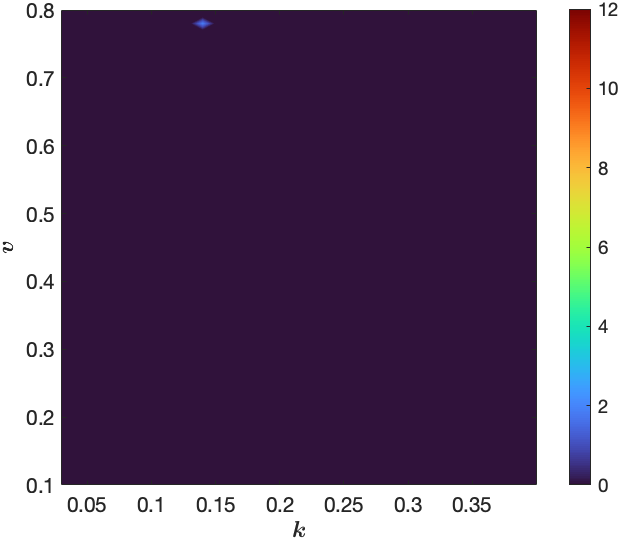}
        \label{fig:fixedE250-k-v}}\hfill
    \subfloat[]
    {\includegraphics[width=0.30\textwidth]
        {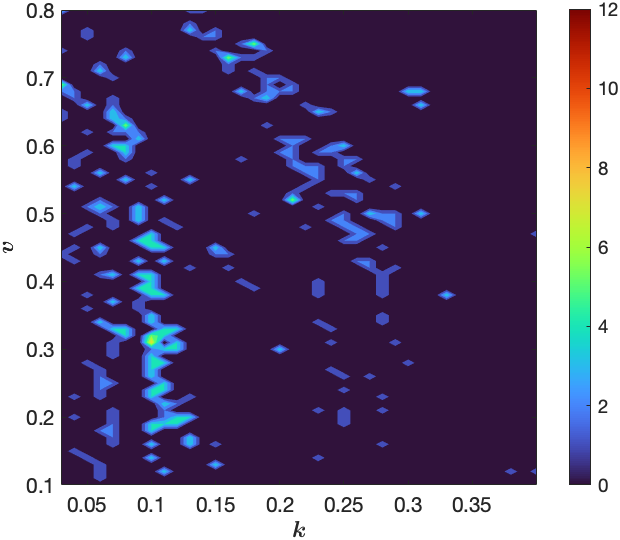}
        \label{fig:fixedE400-k-v}}\hfill
    \subfloat[]
    {\includegraphics[width=0.30\textwidth]
        {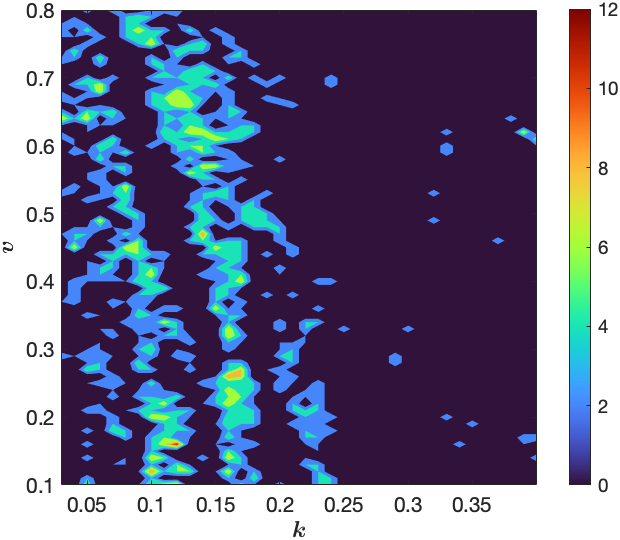}
        \label{fig:fixedE550-k-v}}\hfill
    \caption{\label{fig:fixedE} 
    Kink-antikink pair production at fixed initial energies
    (a)$E_{\psi,0}=250$, (b)$E_{\psi,0}=400$, (c)$E_{\psi,0}=550$  showing
    enhanced production at small $k$ (wider wavepackets) scattering at low velocities. 
    In Fig.~\ref{fig:fixedE250-k-v} there is only one case of kink production at
    $k=0.14$ and $v=0.78$.
}
\end{figure*}

With somewhat different parameters, the evolution can be quite different, with the production
of several kink-antikink pairs. An example is shown in Fig.~\ref{fig:phi-psi-evol-chaotic-main} for the
parameters $v=0.25$ and $A=13.5$. Now the evolution leads to a lot more fluctuations of $\phi$
and there are many zeros of $\phi$ at the final time. With further evolution we expect some of the
zeros to annihilate but by our criteria, described in Sec.~\ref{numerical}, this final state 
contains five kink-antikink pairs. Now the energy density in $\rho$ is more spread out
as in Fig.~\ref{fig:Z-ED-chaotic}
and the total energies in the fields shows an interesting crossover
in Fig.~\ref{fig:total-energy-chaotic}  
where most of the energy ends up in the quantum field $\rho$.

In this case, we start out with a higher initial energy $\sim 750$ but we end up with five
kink-antikink pairs with energy $\sim 10$ which is a higher fraction of the initial energy than in
the case of Fig.~\ref{fig:phi-psi-evol-docile-main}. However, it is not clear how many of
the five kink-antikink pairs will survive at very late times. The complexity is shown in
Fig.~\ref{fig:phi-zeros} where we plot the zeros of $\phi$ as a function of time.
In the case of Fig.~\ref{fig:phi-psi-evol-docile-main}, the zeros are shown in 
Fig.\ref{fig:zeros-docile} and there is only one kink-antikink pair
and they are separating with velocity $\sim \pm 0.68$. In the case corresponding to
Fig.~\ref{fig:phi-psi-evol-chaotic-main}, the plot of zeros of $\phi$ is shown in
Fig.~\ref{fig:zeros-chaotic}. The outermost zeros are moving apart with
$\sim \pm 0.78$ but the inner ones are slower and some annihilations are very
likely. 

In order to find initial conditions that are favorable for the production of kinks we
have evolved the system for the range of initial conditions given in \eqref{Avk}
and checked which initial conditions lead to kink production. Our results 
are shown in Fig.~\ref{fig:kinkSpect} and indicates favorable conditions for
kink production for large $A$ and small $v$ (at least in the $k=0.1, 0.3$ cases). 
However, the results suggest a fractal structure
and there are lots of holes in the parameter space where otherwise one may
expect kink production. There are also isolated special places in parameter space where
a large number of kinks are produced.

From Fig.~\ref{fig:kinkSpect} it is clear that choosing wider Gaussian wavepackets (smaller $k$)
is more favorable to kink production. A first thought is that smaller $k$ might imply higher initial 
energy which would explain the greater rate of kink production but that is not necessarily the case 
since the initial energy in $\psi$ can be calculated explicitly in the limit of large $x_0$
by using \eqref{psit0} and \eqref{dotpsit0} in \eqref{Ephipsi} (see Appendix~\ref{appA}),
\be
E_\psi(t=0) =   \sqrt{\frac{\pi}{2}} \, \gamma A^2 \biggl [ (1+v^2) \sqrt{k} 
+ \frac{m_{\psi}^2}{\sqrt{k}} (1-v^2) \biggr ]
\label{Eboosted}
\ee
For fixed $v$, $E_\psi(t=0)$ is minimum when
\be
k=k_* = m_\psi^2 \left ( \frac{1-v^2}{1+v^2} \right )
\ee
and the energy does not monotonically increase with decreasing $k$.
While it is true that for our choice of values of $k$ and $v$ in \eqref{Avk}, the initial
energy is higher for smaller values of $k$, the energies for $k=0.1$ and $k=0.3$
and with $v=0.8$ are very close, to within $6\%$,
yet there is much more kink production 
with $k=0.03$ than with $k=0.1$. This suggests that a more spread out wavepacket in the initial
conditions is favorable for kink production.

To explore the effect of changing Gaussian width and wavepacket velocity, we
have performed several runs in which the total initial energy in $\psi$ is fixed. We
implement this by choosing 
\be
A^2 = 
\frac{E_i}{ \sqrt{\frac{\pi}{2}} \, \gamma \biggl [ (1+v^2) \sqrt{k} 
+ \frac{m_{\psi}^2}{\sqrt{k}} (1-v^2) \biggr ]}
\label{Asq}
\ee
for some choice of initial energy $E_i$. We scan
over parameters $k$ and $v$, adjusting $A^2$ according to \eqref{Asq} so that
the initial energy stays fixed (up to very tiny corrections due to the quantum
fluctuations of $\rho$ and exponentially small corrections due to the overlap
of the two wavepackets). The
results are shown in Fig.~\ref{fig:fixedE} for fixed initial energy of 250, 400 and
550. The first feature that stands out is that for fixed $E_{\psi,0}=250$ 
there is only a very small area that yields kink production as seen in Fig.~\ref{fig:fixedE250-k-v},
which suggests an energy threshold for kink production for the model. Figs.~\ref{fig:fixedE400-k-v} 
and \ref{fig:fixedE550-k-v} exhibit similar band patterns, although the location and size of these 
bands are slightly different, for example, the gap between the two bright bands is larger for 
$E_{\psi,0}=400$. The plots show the general trend that higher energy, wider wavepackets, 
and slower scattering velocities create favorable conditions for kink production.

\section{Conclusions}
\label{conclusions}

We have studied the creation of classical kinks by scattering classical
wavepackets but where the wavepacket and kink interactions are mediated by
a quantum field. This setup was motivated by the case of monopole production in
light on light scattering, since classical light on light scattering is trivial and only 
becomes non-trivial when quantum effects, such as box diagrams, are included. 
However, there
are differences between our toy model and the physical case of magnetic monopole
production. In the latter, light on light scattering would produce heavy gauge bosons
due to quantum interactions and the heavy gauge bosons themselves would form
the magnetic monopoles. This is unlike in our toy model where we have chosen
a classical field, distinct from the quantum fields, that composes the kinks. Our
choice was necessary because kinks are conveniently described as classical 
configurations, not as a conglomerate of quantum particles.

We have scanned a set of parametrized initial conditions for successful kink
production. Certain trends are clear within our analysis. The initial conditions
that led to kink production in our simulations all have total energy that is 
$10^2-10^3$ times the energy in a kink-antikink pair.
However, the energy per quanta need not be large and is of order $m_\psi$ as
the velocities are only mildly relativistic. In fact, we found that it is somewhat 
favorable to choose moderate velocities, $v \sim 0.5$, but to have large 
values of the amplitude $A$ corresponding to a large number of quanta
in the initial state, $N \sim E/m_\psi \sim 10^2-10^3$.
There is no systematic trend, however, and there are
``holes'' in our scan of parameter space as seen in Fig.~\ref{fig:kinkSpect}.
This suggests that there may be resonances at work -- if certain frequencies
match, kink production is more favorable.
It would be of interest to find initial conditions with less energy and that convert
into kink-antikink pairs more efficiently.

We have investigated the effect of the width 
of the initial wavepackets and the scattering velocity on the kink production with fixed 
initial energies. 
It is clear from Fig.~\ref{fig:fixedE} that there is a lower energy threshold and wider 
wavepackets with lower velocities provide better conditions for kink production. 
We also observe a band structure where certain widths seems more favorable than others.
It's also worth noting that production is significantly lower for smaller widths (higher k) in 
all the cases.

Our analysis is of an exploratory nature as magnetic monopole production
in the real world is much more complicated. Yet our analysis suggests that
scattering at high luminosities is much more desirable than scattering at
high energies if the goal is to produce magnetic monopoles.

\acknowledgements
We thank Mainak Mukhopadhyay and George Zahariade 
for useful comments. TV is grateful to the University of Geneva for hospitality while
this work was being done.
This work was supported by the U.S. Department of Energy, Office of High Energy 
Physics, under Award No.~DE-SC0019470.

\appendix

\section{Energy of the initial wavepackets}
\label{appA}

We ignore the quantum corrections to the energy of the initial wavepackets as these are small
(see Figs.~\ref{fig:total-energy-docile} and \ref{fig:total-energy-chaotic}) and consider widely separated 
wavepackets in which case the initial energy is
just twice that of a single wavepacket,
\be
E_{\psi,0} = 2  \int dx \  \left[ \frac{1}{2} \ (\dot{\psi}^2 + \psi^{'2}) + \frac{m_\psi^2}{2} \psi^2 \right ]
\ee
with 
\be
\psi = A e^{-kX^2}, \ \
{\dot \psi} = \gamma v A (-2kX) e^{-kX^2}
\ee
where $X = \gamma (x+x_0)$. This leaves us with Gaussian integrals and we get,
\be
E_\psi(t=0) =   \sqrt{\frac{\pi}{2}} \, \gamma A^2 \biggl [ (1+v^2) \sqrt{k} 
+ \frac{m_{\psi}^2}{\sqrt{k}} (1-v^2) \biggr ]
\label{Eboostedappendix1}
\ee
This expression can also be written as,
\be
E_\psi(t=0) =   \gamma E_{\psi,v=0} +
\sqrt{\frac{\pi}{2}} \, \gamma v^2 A^2 \biggl [ \sqrt{k}  - \frac{m_{\psi}^2}{\sqrt{k}} \biggr ]
\label{Eboostedappendix}
\ee
Note that $E_\psi(t=0)$ is not $\gamma E_{\psi,v=0}$ as we might expect from 
special relativistic boosts. This is because the Gaussian wavepacket is not a solution of
the static equations of motion. Only static solutions of the equations of motion obey
the special relativistic transformation when boosted.

For $k < m_\psi^2$, the term in square brackets in \eqref{Eboostedappendix} can be
negative and it may happen that the initial energy {\it decreases} with increasing $v$.

\bibstyle{aps}
\bibliography{paper}

\end{document}